# Superconductivity in the vicinity of ferromagnetism in oxygen free perovskite MgCNi$_3$: An experimental and DFT (Density Functional Theory) study

Anuj Kumar[1,2], Rajveer Jha[1], Shiva Kumar[1], Jagdish Kumar[1,3], P. K. Ahluwalia[3], R. P. Tandon[2] and V.P.S. Awana[1,*]

[1]*Quantum Phenomena and Applications, National Physical Laboratory (CSIR), New Delhi-110012, India*

[2]*Department of Physics and Astrophysics, University of Delhi, North Campus, New Delhi-110007, India*

[3]*Department of Physics, Himachal Pradesh University, Summerhill, Shimla-171005, India*

We report synthesis, structural, magnetic, specific heat and Density Functional Theory (DFT) studies on MgCNi$_3$ superconductor. Polycrystalline MgCNi$_3$ samples are synthesized through standard solid state reaction route and found to crystallize in cubic perovskite structure with space group *Pm*3*m*, without any detectable trace of Ni impurity. Both *AC* and *DC* magnetization exhibited superconducting transition ($T_c$) at around 7.25 K. The lower critical field ($H_{c1}$) and irreversibility field ($H_{irr}$) are around 140 Oe and 11 kOe respectively at 2 K. The upper critical field ($H_{c2}$) being determined from in-field *AC* susceptibility measurements is 11.6 kOe and 91.70 kOe with 50% and 90% diamagnetism criteria respectively. Heat capacity ($C_p$) measurements are carried out under applied field of up to 140 kOe and down to 2 K. The Sommerfeld constant ($\gamma$) and Debye temperature ($\Theta_D$) as determined from low temperature fitting of $C_p(T)$ data to Sommerfeld-Debye model are 36.13 mJ/mole-K$^2$ and 263.13 K respectively. The Bardeen-Cooper-Schrieffer (BCS) parameter ($2\Delta/K_BT_c$) is around 3.62, suggesting MgCNi$_3$ to be an intermediate coupled *BCS* superconductor with value $\lambda$ = 0.69. Although the density functional theory (DFT) calculations exhibited the compound to be non-magnetic but with spin fluctuations, the experimental isothermal magnetization *MH* loops at 20, 50, 100, 200 and 300 K showed some ferromagnetic nature in this compound with coercive field ($H_c$) of around 50 Oe at 20 K. The Ni$^{3d}$ states play dominant roles near the Fermi levels and there is strong hybridization between Ni$^{3d}$ and C$^{2p}$ states. It seems that MgCNi$_3$ is superconducting in close proximity of ferromagnetism.

# I. Introduction

The discovery of superconductivity in the MgB$_2$ compound with $T_c$ ~ 39 K [1] has created an interest for the search of new inter-metallic compounds with high superconducting transition temperatures. Later, superconductivity at ~ 8 K was discovered in the inter-metallic compound MgCNi$_3$ [2]. Instead of layered MgB$_2$ (space group *P6/mmm*), where the Boron layers are alternately stacked on the Mg layers, the MgCNi$_3$ has the classical cubic perovskite structure (space group *Pm3m*), resembling the high $T_c$ cuprates. Despite the strong expectation for ferromagnetism due to full occupancy of Ni, in MgCNi$_3$, it shows superconductivity but not the ferromagnetism [2]. MgCNi$_3$ is the only known non-oxide perovskite that exhibits superconductivity with a critical temperature of ~ 8 K [2]. Because of the large content of Ni in MgCNi$_3$, this material lies in close proximity to ferromagnetism [3]. The occurrence of superconductivity in MgCNi$_3$ is unusual and a border line case between the conventional non-magnetic and ferromagnetic superconductors [4]. The conduction electrons of MgCNi$_3$ are derived predominantly from Ni [3, 4]. Carbon atom in MgCNi$_3$ plays a critical role in its superconductivity. Single-phase superconducting compound (MgC$_x$Ni$_3$) is formed in a range of Carbon content (1.3 <$x$< 1.6) [5, 6]. In brief, MgCNi$_3$ is a superconductor equipped with its rich physics and hence of tremendous interest to both experimentalists and theoreticians [2-5, 7].

MgCNi$_3$ forms in a three-dimensional perovskite structure. Comparison to a familiar oxide perovskite such as CaTiO$_3$, for example, indicates the structural equivalencies between Ca and Mg, Ti and C, and O and Ni. Instead of a Ti atom, a C atom is located in the body-centered position surrounded by a Ni (6)-octahedra cage. Such a large amount of nickel in MgCNi$_3$ makes it an analog of the three-dimensional layered nickel borocarbides, typified by LuNi$_2$B$_2$C with a $T_c$ near 16 K [8]. In general, the co-ordination of Carbon with other cations in structural alloys or compounds could give rise to various innovative physical properties. Often the *3d* magnetic atoms when coordinated with Carbon do strip off their moments. For example Ni being coordinated with Carbon in MgCNi$_3$ [3-4, 7] and LaNiC$_2$ [9] superconductors possesses nearly no magnetic moment. The observance of superconductivity in Ni based compounds is not unprecedented and already reported for the quaternary borocarbides (RENi$_2$B$_2$C) [7], MgCNi$_3$ [3-4, 7], LaNiC$_2$ [9] and binary alloy of Bi and Ni (Bi$_3$Ni) [10, 11] etc. One of the moot questions in regards to MgCNi$_3$ superconductor is about the nature of its magnetism. In particular, it is difficult to get the end compound without the traces of Ni impurity [8]. Small amount of un-reacted Ni, being magnetic and even ferromagnetic (*FM*) with Curie temperature above 600 K, could mess up the net outcome in magnetization measurements. Often, this compound is synthesized with nominal excess C, i.e., $MgC_xNi_3$ [2, 6, 12-14], and the optimization of stoichiometric $MgCNi_3$ is scant. The theoretical calculations [7] show that in $MgC_xNi_3$ (x=0.8-1.0), the proportionality of magnetism increases with decreasing x.

As far as the physical properties are concerned, the measurements of thermopower indicated that the carriers in $MgCNi_3$ are electron type [15, 16]. The thermal conductivity of the $MgCNi_3$ is of the order as for inter-metallics, but larger than that of borocarbides and smaller than $MgB_2$ [16]. The specific heat [2, 12, 17] and resistivity [2, 18, 19] measurements on bulk polycrystalline samples implied a moderately coupled superconductor [17, 19]. The penetration depth measurements distinctly showed a non-*s*-wave BCS feature at low temperature [20]. Lately, detailed penetration depth and specific heat analysis on $MgCNi_3$ on single crystals confirmed *s-wave* superconductivity with fully open gap and strong-coupling [22, 23]. On the other hand the Hamiltonian-based model calculations suggest $MgCNi_3$ to be a *d*-wave superconductor [21]. Recently, a $C^{13}$ isotope effect is reported with $\alpha_C = 0.54$ (3), which indicates that $MgCNi_3$ is predominately a phonon-mediated superconductor and confirms the important role of Carbon in superconductivity [24]. Most of scattered physical properties and theoretical calculation of $MgCNi_3$ are reviewed in ref. [6]. It appears that there exist some contradictory reports regarding physical property measurements on bulk and single crystalline $MgCNi_3$ [2, 6, 12-23]. The recent discovery of the iron-pnictide high $T_c$ superconducting materials [25, 26] opened a new window for condensed matter community. Many of these iron-pnictide materials contains two magnetic elements (Fe and rare-earth metal) and the interplay between magnetic order and superconductivity is very important and interesting [26].

Here we report the synthesis temperature optimization and host of superconducting properties for the stoichiometric $Mg_{1.2}C_{1.6}Ni_3$. We optimized the synthesis temperature for $Mg_{1.2}C_{1.6}Ni_3$ and achieved a single phase compound without any detectable impurity of Ni within the limit of X-ray powder diffraction. Excess amount of Mg is taken due to its volatile nature. All the superconducting characteristics of the compound are reproduced. Detailed heat capacity ($C_p$) measurements are carried out under applied field of up to 140 kOe and down to 2

K. The Sommerfeld constant ($\gamma$) and Debye temperature ($\Theta_D$) are 36.13mJ/mole-K$^2$ and 263.13 K respectively. The *BCS* parameter ($2\Delta/K_BT_c$) is around 3.62, and electron-phonon coupling constant is $\lambda = 0.69$, suggesting MgCNi$_3$ to be an intermediate coupled *BCS* superconductor. It is found that this compound possesses *some FM correlations* along with the bulk superconductivity at below 7.25 K.

## II. Experimental details

Various samples of Mg$_{1.2}$C$_{1.6}$Ni$_3$ are synthesized through standard solid state reaction route. The starting materials are Mg flakes (99% Riedal-de Haen), Ni (99.99% Sigma Aldrich) and amorphous Carbon powder. The excess amount of Mg and C in the synthesis is taken to compensate their volatile deficiency during heat treatments [2, 27]. The stoichiometric amounts of high purity ingredients (Mg, amorphous C and Ni) were ground thoroughly and palletized using hydraulic press. These pallets were putted in iron capsule and then heated at 600°C for 2 hours in the continuous flow of argon gas with purity (99.9%), followed by heating at higher temperatures i.e. 900, 930, 950, 975 & 1000°C for 3 hours and finally cooled down to room temperature in the same atmosphere. Following the final heat treatment, the samples are analyzed by powder X-ray diffraction using Cu K$_\alpha$ ($\lambda$ = 1.54Å) radiation. The Magnetic (*DC*, *AC* susceptibility and isothermal magnetization) and Heat Capacity [$C_P(T)H$] measurements are carried out on *PPMS* (physical property measurement system) a Quantum Design-USA down to 2 K and varying fields of up to 140 kOe. *DFT* calculations are performed using density functional based approach employing Full Potential Linear Augmented Plane Wave (*FP-LAPW*) basis as implemented in *ELK* code [7]. The Brillouin zone integrations are performed using 15x15x15 mesh containing 120 k-points. The values of Muffin tin radii for Mg, C and Ni were 2.00, 1.45 and 2.00 *a.u.* (atomic units) respectively. We used generalized gradient approximation [28] for solids (*GGA PBEsol*) to describe the exchange and correlation for the system. The lattice constant for MgCNi$_3$ is determined by plotting total energy versus lattice constant and was found to be 7.12 *a.u.* which is within 1.1% of experimental value 7.20 *a.u.*

## III. Results and Discussion

Fig.1. depicts the room temperature X-ray diffraction (XRD) patterns of studied $Mg_{1.2}C_{1.6}Ni_3$ processed at different temperatures. For clarity the XRD patterns of 930°C, 950°C and 975°C processed samples are shown. It is clear that though the 930°C and 975°C processed samples are contaminated with Ni impurity (marked with *), the 950°C sample is phase pure. The 950°C processed sample is crystallized in cubic oxygen free perovskite structure with space group *Pm3m*. The X-ray diffraction data was fitted using Rietveld refinement (*FullProf* Version). The coordinate positions for the atoms are found as: Mg: 1*a* (0, 0, 0), C: 1*b* (0.5, 0.5, 0.5) and Ni: 3*c* (0, 0.5, 0.5). It is clear that 950°C is the optimum sintering temperature for $MgCNi_3$. Lattice parameters of the studied $Mg_{1.2}C_{1.6}Ni_3$ (sintered at 950°C) are found to be 3.796 (6) Å. Worth mentioning, is the fact that even a slight change in sintering temperature of say 10-20°C results in Ni impurity. After various trials we optimized the sintering temperature of phase pure $Mg_{1.2}C_{1.6}Ni_3$ to be 950°C. The lattice parameters are in good agreement with earlier reports for nominal $MgC_{1.6}Ni_3$ [2, 6].

Fig. 2(a) shows the DC magnetization of the phase pure $Mg_{1.2}C_{1.6}Ni_3$ sample (sintered at 950°C) in *FC* and *ZFC* situations at 10 Oe field. Sample shows bulk superconductivity with an onset temperature ($T_c^{onset}$) of 7.25 K. Fig. 2(b) and 2(c) depicts the imaginary and real part of *AC* magnetization for the same. $T_c^{onset}$ of 7.25 K is observed in both *DC* and *AC* magnetization. There seems to be large irreversibility in DC magnetization plot. As will be discussed in next figure related to AC magnetization results, this compound seemingly do have no grain boundary contributions. This explains its single intra grain transition with possibility of large irreversibility field. Fig. 3 represents the *AC* magnetization of the same sample, which also confirms the bulk superconductivity at around 7.25 K in the present compound. Further, *AC* magnetic susceptibility measurements at 333 *Hz* and varying amplitude of 3 to 11 Oe (Figure 3) showed nearly no change in the superconducting transition. It seems that the superconducting grains are well coupled in this compound and there is nearly no grain boundary contribution in the superconductivity of $MgCNi_3$.

The low field (< 1000Oe) isothermal *MH* plots of studied $MgCNi_3$ at *T* = 2, 4 and 6 K are shown in Fig. 4. To determine the lower critical field ($H_{c1}$), we draw a tangent line from the origin on the curve. $H_{c1}$ is defined as deviation point between the tangent line and the curve at 2

K. The $H_{c1}$ is around 140 Oe at 2 K and decreases monotonically with increase in temperature to 85 and 35 Oe at 4 and 6 K respectively. Isothermal magnetization (*MH*) plots at 2 K in high field range of up to 120 kOe in four quadrants for MgCNi$_3$ are shown in Fig. 5. The lower critical field ($H_{c1}$) and irreversibility field ($H_{irr}$) are around 140 Oe and 11 kOe respectively at 2 K. Fig. 6 shows the upper critical field ($H_{c2}$) being determined from in-field *AC* susceptibility measurements, which is 11.6 kOe and 91.70 kOe with 50% and 90% diamagnetism criteria respectively. It is clear that irreversibility field ($H_{irr}$) and upper critical field ($H_{c2}$) both are of the same order of magnitude. The isothermal magnetization *MH* plot at 20 K in low field range of <1000 Oe are shown in Fig. 7. The ferromagnetic nature of *MH* plot at 20 K indicates that superconductivity of MgCNi$_3$ lies in the vicinity of ferromagnetism. Theoretical calculations predicts about non-magnetic nature of MgCNi$_3$, to be discussed in next sections. There is a possibility that ferromagnetism could be due to small impurity of Ni, although it is not detectable in our X-ray diffraction limits, and will be ruled out in next sections. The Z*FC* and *FC* magnetization of the compound in temperature range of 300 K to 5 K done under 1000 Oe applied field is shown in inset (a) of Fig. 7. It seems that though the magnetization behaviour is typical of a ferro-magnet, the moment value is quite small for which Ni might be the possible culprit. However, careful look of the *MH* plots (inset (b) Fig.7), show that the saturation moment is increasing even till 50 kOe. In case of Ni, the FM moment saturates completely below 10 kOe [29] hence the presently observed non saturating ferromagnetic correlations in MgCNi$_3$ are intrinsic and not driven by Ni impurity. This rule out the possibility of the presence of undetected Ni within XRD limits of less than 1% being contributing to the matrix para-magnetism of MgCNi$_3$. In fact in experimental literature yet there are no data on normal state (above Curie temperature, T$_C$) magnetism of this important superconductor. Our data is complete in terms of various physical properties of MgCNi$_3$ and is quite clean with in XRD detection limit (Fig.1). Further our results are thought provoking as far as the possibility of ferromagnetism is concerned in MgCNi$_3$ along with superconductivity.

We have performed density functional theory (DFT) calculations for the mentioned compound using Full Potential Linear Augmented Plane Wave method (*FP-LAPW*) as implemented in *ELK* code. The numerical convergence of different parameters was ensured. We have used generalized gradient approximation (*GGA*) to perform calculations. The calculated *DOS* was found degenerate for both the spins. The results are shown in Fig. 8. The atom resolved

*DOS* shows that most of the states near Fermi level arise from Ni *d*-states and Mg and C states are relatively small. The bonding Mg *s*-states (not shown here) are distributed from 0.1 to 0.26 Hartree below Fermi-level that shows covalent bonding with Ni *d*-orbital, whereas corresponding anti-bonding orbitals are found from 0.04 to 0.4 Hartree above Fermi level. The value of *DOS* at Fermi-level is found to be 3.76 states/eV-cell. The sharp peak seen just below Fermi level corresponds to well known van Hove singularity (*vHs*) for this system which is also seen in certain high temperature superconductors (*HTSc*). The value of *DOS* at Fermi-level is found to be 102.50 states/Hartree-Cell (3.76 states/ eV-Cell). It is interesting to note that there are many reports which show variations in the calculated values depending upon method of calculation used. For example, Dugdale and Jarlborg [30] report $N(E_F)$ to be 6.35 and 3.49 states/ eV-cell, P. Jiji et.al. 5.35 states/ eV-cell [7], Mazin and Singh report as 4.99 states/ eV-cell [31], Shim *et al.* as 5.34 states/ eV-cell [32] and Rosner *et al.* as 4.8 states/ eV-cell [3]. Our calculated value of 3.76 states/ eV-Cell lies, within reported values [3, 7, 31, 33].

We found that for perfectly stoichiometric MgCNi$_3$ compound there was no magnetic moment on Ni atom and self consistent calculations resulted into a nonmagnetic solution. To check any possibility of a local minimum around zero moment state, we did fixed spin moment (FSM) calculations, which shows that it costs extra energy to induce any magnetic moment into the system. The results of FSM calculations are shown in inset of Fig. 8. This, rules out the possibility of ferromagnetism in perfectly stoichiometric MgCNi$_3$. Calculations done on hypothetical compound MgNi$_3$ using same lattice parameters however resulted into a ferromagnetic solution. Thus role of Carbon seems important for magnetism. Also there are calculations on Carbon deficient MgCNi$_3$ [7] using coherent potential approximation (*CPA*) that shows that magnetism cannot appear even up to very high degree of deficiency of Carbon. In fact we did spin polarised calculations using *SPRKKR* code that uses *CPA* and found that Carbon concentration of up to 30 percent is enough to kill whole magnetism of this compound. The loss of magnetism in this compound can be attributed to strong hybridisation of Ni-*3d* bands with C-*2p* bands. This can also be seen from the *DOS* where C-*2p* and Ni-*3d* states overlap over a large energy range. Such strong hybridisation results into delocalisation of Ni-*d* states thus lowering itinerant character of the material. To check the possibility of observed ferromagnetism, we have also replaced Mg by Ni up to 10 percent using *SPRKKR* code, which also did not result into any ferromagnetic solution. Thus possibility of small replacement of Mg by Ni resulting into magnetism can be overruled.

To check the validity of *CPA* we have done spin polarised calculations on $MgC_{0.5}Ni_3$ using supercell approach with *ELK* code. We used 2×1×1 supercell with one C atom replaced by a vacancy. The self consistent calculations done on this supercell resulted into a ferromagnetic ground state. The detailed origin of observed ferromagnetism is as follows. Each Ni atom in $MgCNi_3$ has two Carbon atoms as nearest neighbours. In the mentioned supercell we had six Ni atoms out of which two Ni atoms have both nearest neighbour Carbon atoms present, other two have only one and the rest two have no Carbon atom at their nearest neighbour position. Interestingly there was no magnetic moment on Ni atoms which are bonded to even a single Carbon atom. Only two Ni atoms that are not bonded to any Carbon atom showed a local moment that resulted into a ferromagnetic solution. This is due to absence of any hybridisation of Ni-*d* states with C-*p* states that results into less delocalised Ni-*d* electrons, which may contribute to magnetism. Thus we can here conclude that the observed intrinsic ferromagnetism in the mentioned samples could have come from localised Carbon deficiencies.

We also performed the heat capacity versus temperature measurements at different values of applied magnetic field. Fig. 9 shows the heat capacity plot with applied magnetic field up to 140 kOe. The $T_c$ is to be obtained 7.2 K, through entropy balance of superconducting and normal state [12]. This is in good agreement with $T_c$ (7.25 K) obtained through magnetization data (Fig. 2). The low temperature normal state heat capacity data is fitted using Sommerfeld-Debye model. The low temperature $T^3$ approximation is used for Debye term. From the parameters of fitting we estimated the values of Debye temperature $\Theta_D$ and Sommerfeld constant $\gamma$. The values of $\Theta_D$ and $\gamma$ are found to be 263.13 K and 36.13 mj/mole-$K^2$ respectively [see inset (a) of Fig. 9]. These values are in good agreement with earlier reports [2, 19].

From the value of Sommerfeld constant we have calculated the electronic Density of states at Fermi level $N(E_F)$ using formula.

$$N(E_F) = 3\gamma/\pi^2 K^2_B \qquad (1)$$

There are contradicting reports regarding nature of electron-phonon coupling parameter $\lambda$ and the nature of superconductivity [2, 12, 17-21] in $MgCNi_3$. It is found that $MgCNi_3$ possesses a *BCS*-like *Cp (T)* in the superconducting state [19]. The electron–phonon coupling constant $\lambda$ is ~ 0.66–0.84 [12, 31] as being estimated from the Eliashberg relation [34]

$$\Delta C/\gamma T_c = 1.43 + 0.924\lambda^2 - 0.195\lambda^3 \quad (2)$$

From our specific heat data fitting, we extracted the value of α to be 1.87 [Fig. 10] using modified *BCS* formula [35]. The specific heat jump at $T_c$ ($\Delta C/\gamma T c$) is obtained to be 1.81. Using this, the value of coupling constant $\lambda$ is obtained 0.69. This indicates moderate electron–phonon coupling in MgCNi$_3$. On the other hand, the value of $\lambda$ extracted using Equation (3) from ref. [34]

$$N_{expt}(E_F)/N_{calc}(E_F) = 1 + \lambda \quad (3)$$

is 3.02, which is nearly four times in comparison to the one being calculated (0.69) from Eliashberg equation.

The calculated *DOS* from *DFT* calculations and from our heat capacity data are 3.76 states/eV-formula unit and 15.31 states/eV-formula unit respectively. From our calculation based on ref. [34], the value of the electron-phonon coupling parameter $\lambda$ is 3.02 i.e., $\lambda \gg 1$. This shows that electron-phonon coupling in studied compound is strong one. Even if we consider the maximum *DOS* calculated from earlier reports (~5 states/eV-formula unit), the value of $\lambda$ comes out around 2. Considering these values of $\lambda$, the calculated value of $T_c$ with $\Theta_D$ = 263.13 K is around 37.38 K. It seems there are some parameters contributing in *DOS*, which we are lacking in calculations. There are reports regarding two phases of MgCNi$_3$ *α*-phase (Carbon depleted phase) and *β*-phase (Carbon occupancy is 0.96). Both *α*- phase and *β*-phases share the same cubic space group, but *α*- phase lattice parameter is 1.3% smaller and, unlike the *β*-phase, and is not superconducting. The calculations show that in case of *α*- phase, the specific heat $\gamma$ is 50% lower than in the *β*- phase [35]. In case of presently studied sample the possibility of occurrence of *α*- phase is excluded, since we have taken Carbon in access (1.6) and also the sample is superconducting. Moreover, in view of the rather large Stoner parameter in MgCNi$_3$ [31], one may have to take into account the spin fluctuation effect on the superconducting property. Shim et al. [32] used the modified $T_c$ formula including the spin-fluctuation parameter $m_{sp}$ in the McMillan's form [37]. They considered small $m_{sp}$= 0.1, and obtained much reduced $T_c$ = 11.5 K even in the case of $\Theta_D$ = 300 K. Spin-fluctuation is responsible for that higher value (3.02) of $\lambda$, due to excess amount of Carbon [38]. The difference in experimental and fitted electronic specific heat data ($C_{es}/\gamma T$) [see Fig. 10] in lower temperature region indicates about multi band superconductivity. In fact two-band model has already been applied [12, 39] to reconcile the controversies in explaining the $C_p(T)$ of MgCNi$_3$. In any case the contrasting values of λ being obtained through fitting of experimental data (Eliashberg relation) and through calculated *DOS* and experimental *DOS* indicates about mysterious nature of superconductivity in MgCNi$_3$ in presence of strong spin fluctuation effects. Inset (b) of Fig. 9 shows the $C_p(T)$ vs. $T^2$ behaviour of MgCNi$_3$ under applied fields of up to 140 kOe. Bulk $T_c$ is seen as a hump at 7.2 K, which decreases to lower temperatures with increase in field. $T_c$ is seen at 3 K for 100 kOe applied field. This means the upper critical field of the compound is above 100 kOe at 3 K. This value is either comparative or even larger than earlier reports [6]. This shows that the studied MgCNi$_3$ sample is fairly good one.

## IV. Conclusion

Polycrystalline MgCNi$_3$ samples are synthesized and found to crystallize in cubic structure with superconducting transition ($T_c$) at around 7.25 K. *AC* magnetic susceptibility measurements at 333 Hz and in varying amplitude of 3 to 11Oe showed that the superconducting grains are well coupled. The lower critical field ($H_{c1}$) and irreversibility field ($H_{irr}$) are around 140 Oe and 10 kOe respectively at 2 K. The upper critical field ($H_{c2}$) are determined from in-field *AC* susceptibility measurements. Heat capacity ($C_p$) measurements are carried out under applied field of up to 140 kOe and down to 2 K. The Sommerfeld constant (*γ*) and Debye temperature ($\Theta_D$) as determined through low temperature fitting of $C_p(T)$ data. The *BCS* parameter ($2\Delta/K_BT_c$) suggests MgCNi$_3$ to be an intermediate coupled *BCS* superconductor. The contrasting values of λ obtained through fitting of experimental data (Eliashberg relation) and through calculated DOS and experimental DOS indicates about mysterious nature of superconductivity in MgCNi$_3$. Although the density functional theory (*DFT*) calculations shows the perfectly stoichiometric compound to be nonmagnetic, and localised Carbon deficiencies can lead to the formation of *intrinsic ferromagnetic domains along with bulk superconductivity*. The experimental isothermal magnetization (*MH*) measurements showed some ferromagnetic correlations in this compound with coercive field ($H_c$) of around 50 Oe at 20 K. It seems the $MgCNi_3$ is superconducting in close proximity of ferromagnetism. The detailed microscopic nature of observed ferromagnetism however will be interesting to investigate through spectroscopic techniques in future.

## Acknowledgements

The authors would like to thank Prof. R. C. Budhani (Director NPL) for his constant support and encouragement. Author Anuj Kumar would like to thanks *CSIR*, for providing financial support through Senior Research Fellowship (SRF) to pursue his Ph.D. Authors Shiva Kumar and Jagdish Kumar also thanks to CSIR for providing financial support. We acknowledge Dr. Jiji Pulikkotil for many fruitful discussions and suggestions.

# Figure Captions

**Fig. 1**: Rietveld fitted XRD pattern of $Mg_{1.2}C_{1.6}Ni_3$ samples with space group *Pm3m* synthesized in various heating conditions.

**Fig. 2**: (a) and (b) Real and imaginary part of *AC* susceptibility, mesure at. 333 Hz and 7 Oe of $MgCNi_3$ respectively, (c) *DC* magnetic susceptibility *M(T)* in *ZFC* (Zero-Field-Cooled) and *FC* (Field-Cooled) situations at 10 Oe of $MgCNi_3$.

**Fig. 3**: *AC* magnetic susceptibility in both real (*M′*) and imaginary (*M″*) situations at fixed frequency of 333 Hz and varying Amplitudes of 3-11Oe for $MgCNi_3$.

**Fig. 4**: Isothermal magnetization (*MH*) plots at 2, 4 and 6 K in low field range of <1000 Oe for $MgCNi_3$, the lower critical field ($H_{c1}$) is marked.

**Fig. 5**: Isothermal magnetization (*MH*) plots at 2K in high field range of up to 120 kOe in four quadrants for $MgCNi_3$. Inset shows expanded *MH* plots at 2K in low field range of up to 30 kOe in four quadrants for $MgCNi_3$, the $H_{irr}$ is marked.

**Fig. 6**: Isothermal magnetization (*MH*) for real part of *AC* susceptibility (*M′*) with applied field of up to 140 kOe at 2 K for $MgCNi_3$, the upper critical field ($H_{c2}$) is marked.

**Fig. 7**: Isothermal magnetization (*MH*) plot at 20 K in low field range of <1000Oe which indicates about ferromagnetic nature. Inset (a) shows DC magnetization in Zero-Field-Cooled (ZFC) and Field-Cooled (FC) situation at 1000 Oe applied field. Inset (b) shows isothermal magnetization (*MH*) plots at 20, 50, 100, 200 and 300K in the field range of <50 kOe for $MgCNi_3$.

**Fig. 8**: Electronic Density of States for $MgCNi_3$ and contribution of Ni and C states. Inset shows results of fixed spin moment (FSM) calculations done on stoichiometric $MgCNi_3$.

**Fig. 9**: Heat Capacity vs Temperature $Cp$ (*T*) plot of $MgCNi_3$ in temperature range of 2-250 K without applied magnetic field. Inset 'a' shows linear fitting of $Cp/T$ vs $T^2$ – 140 kOe curve (normal state). Inset 'b' shows $Cp/T$ vs $T^2$ plots in magnetic fields 0 to 140 kOe.

**Fig. 10**: Single gap fitting of electronic specific heat below transition Temperature ($T_c$) using α-Model.

**Figure 1**

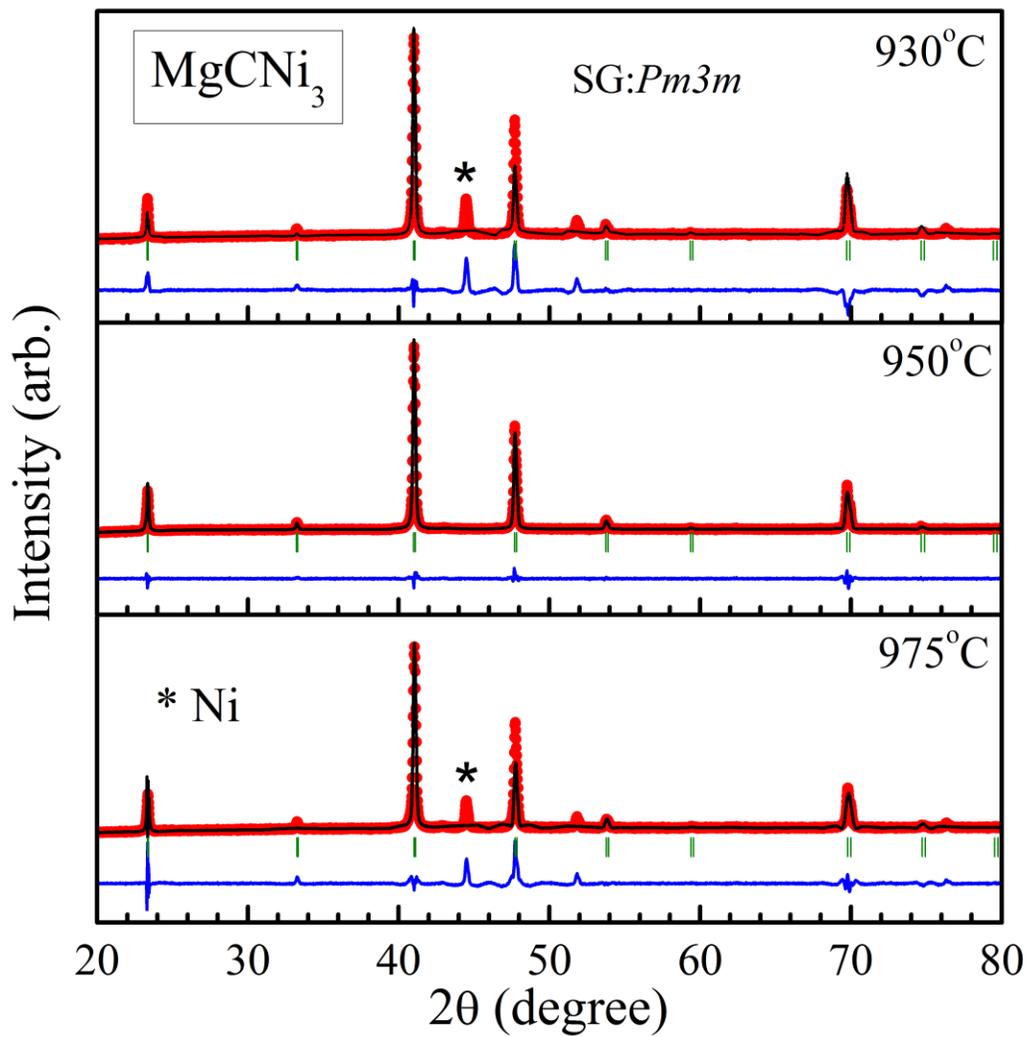

**Figure 2**

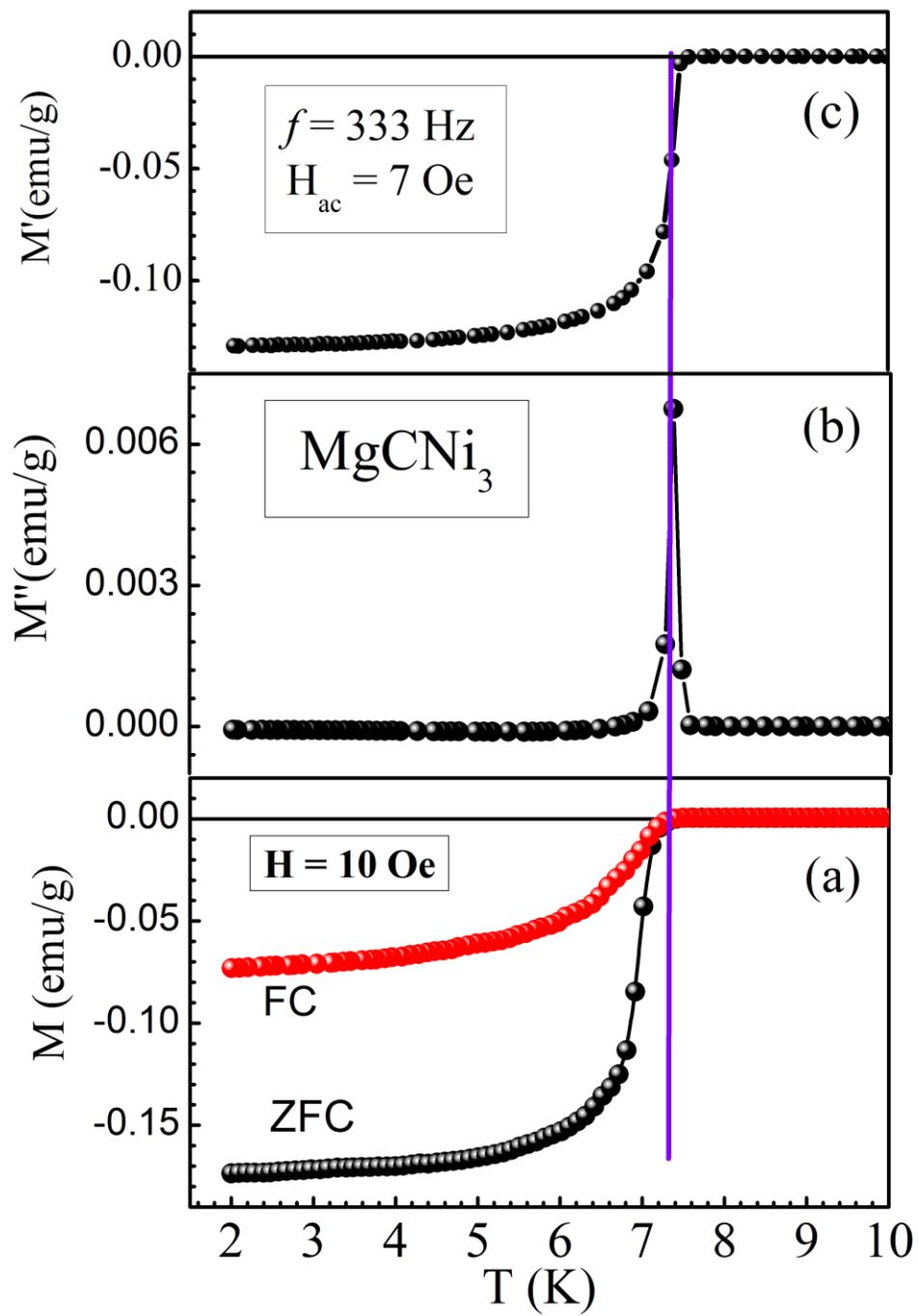

**Figure 3**

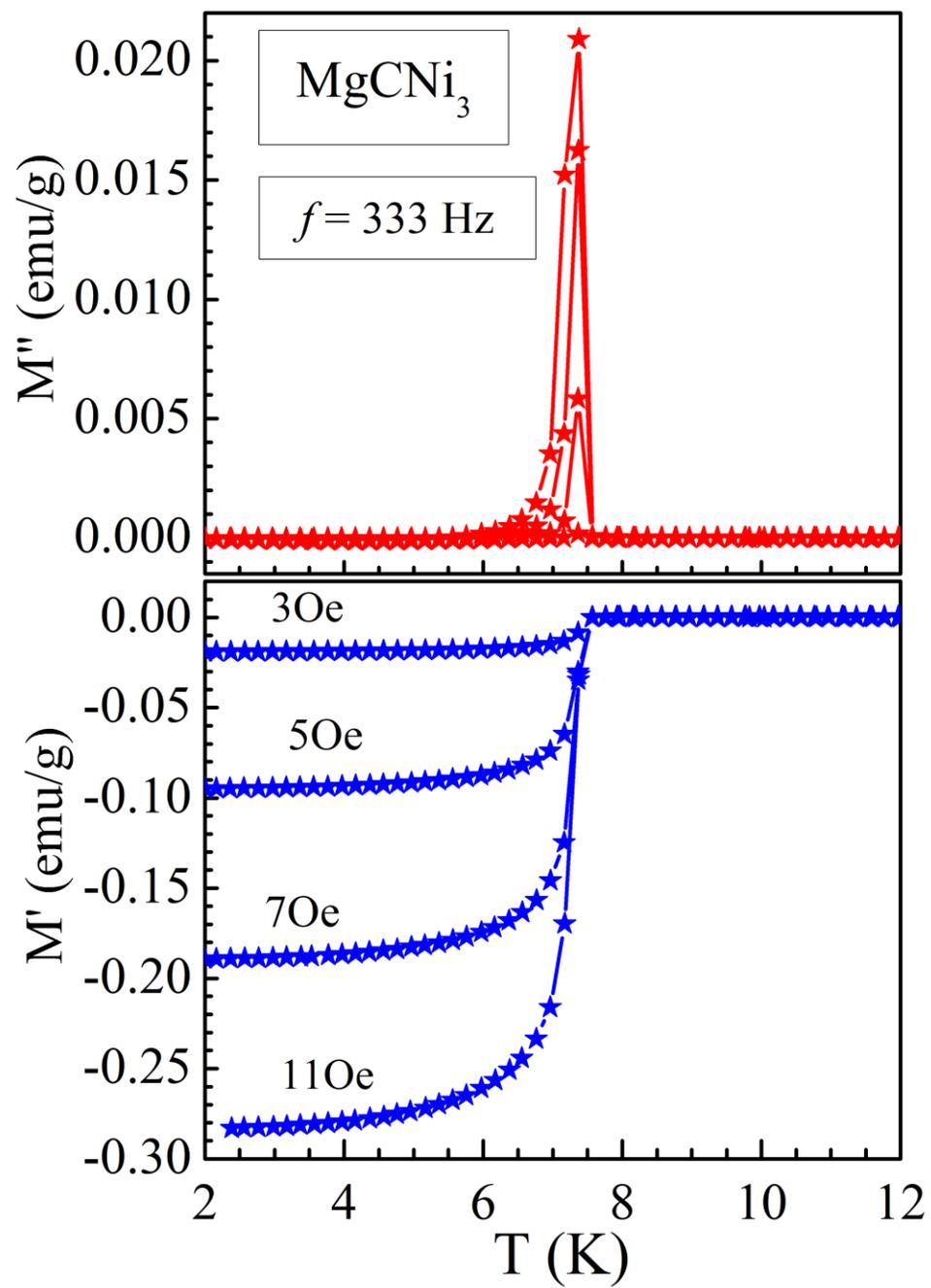

**Figure 4**

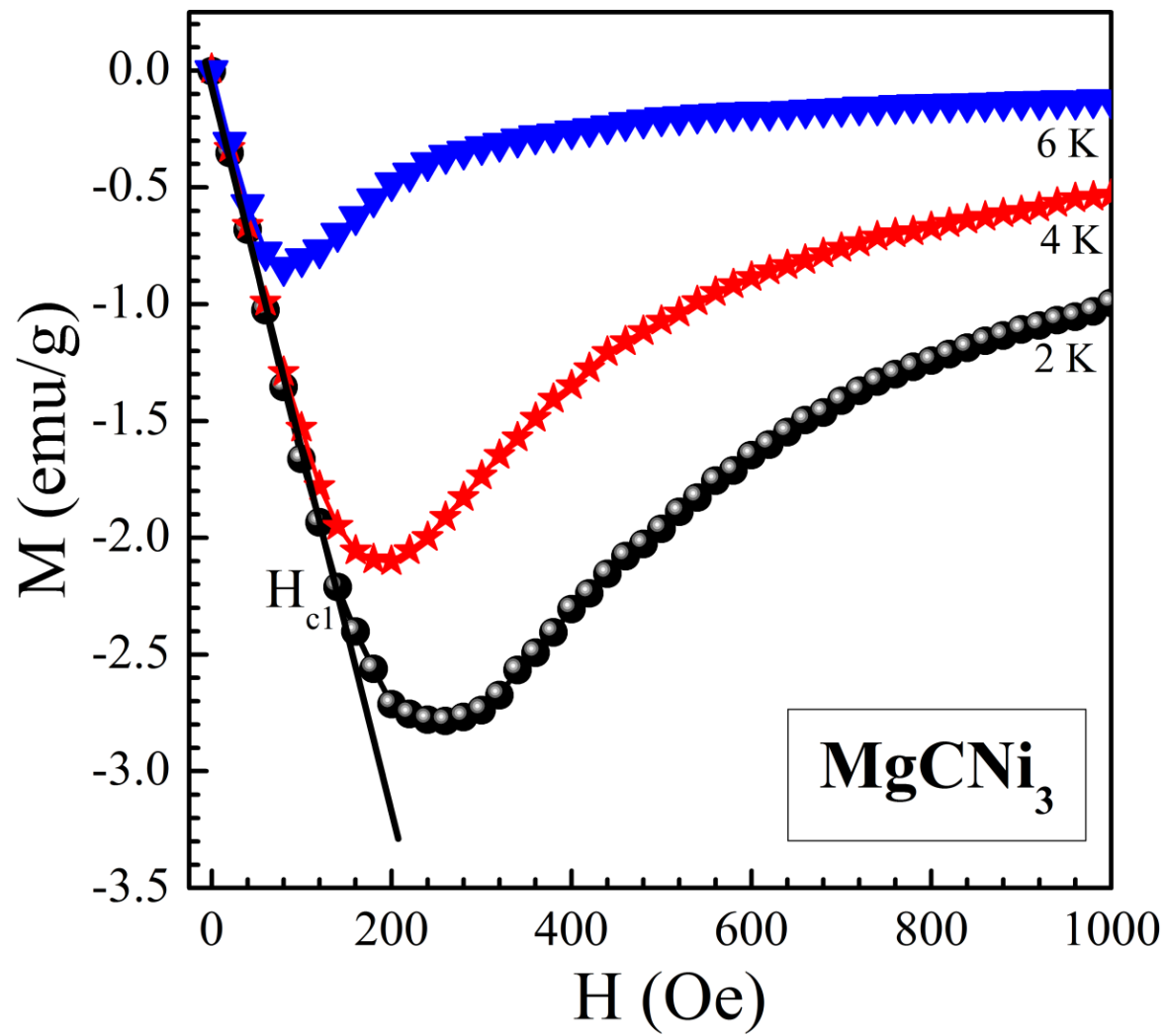

**Figure 5**

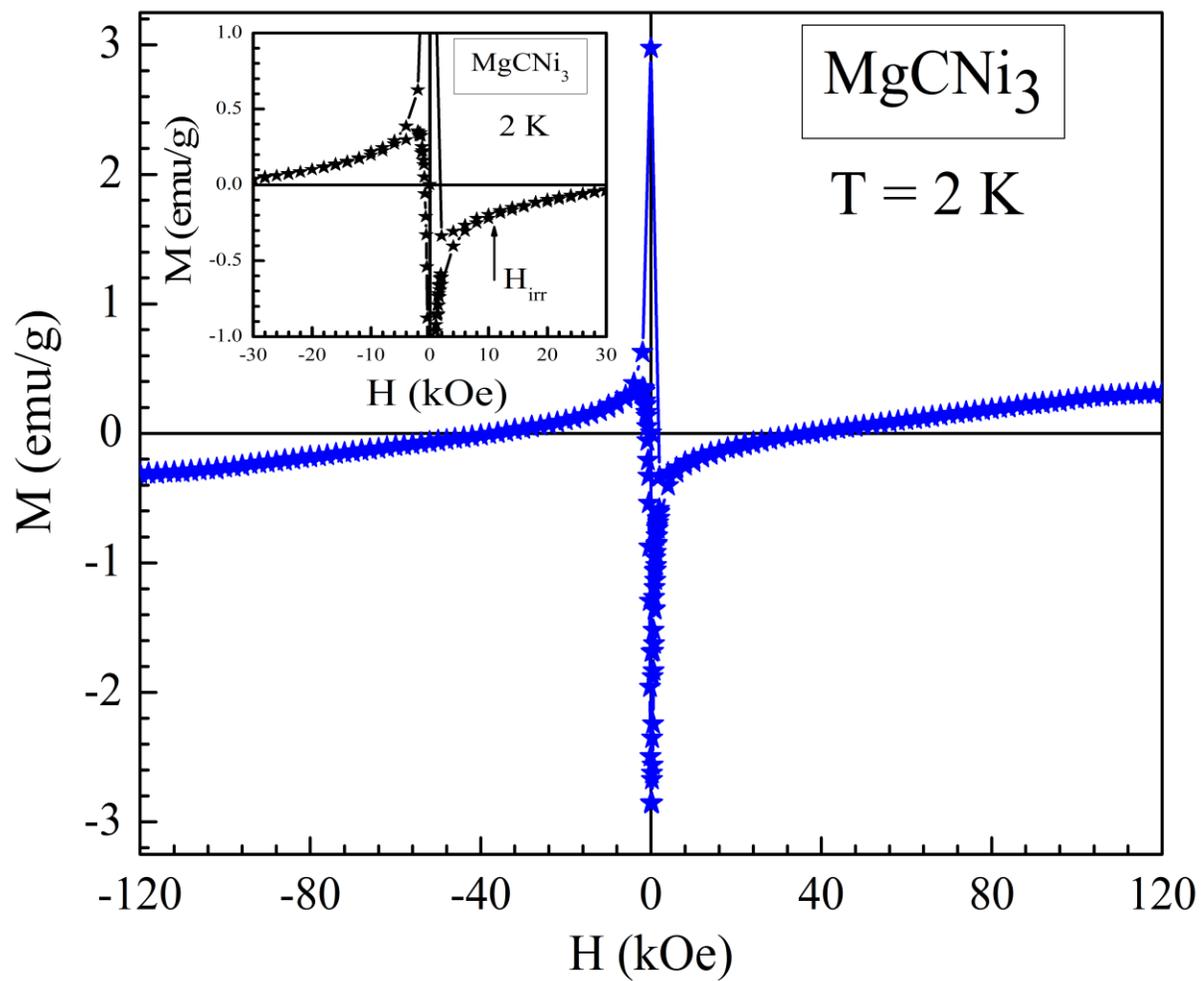

**Figure 6**

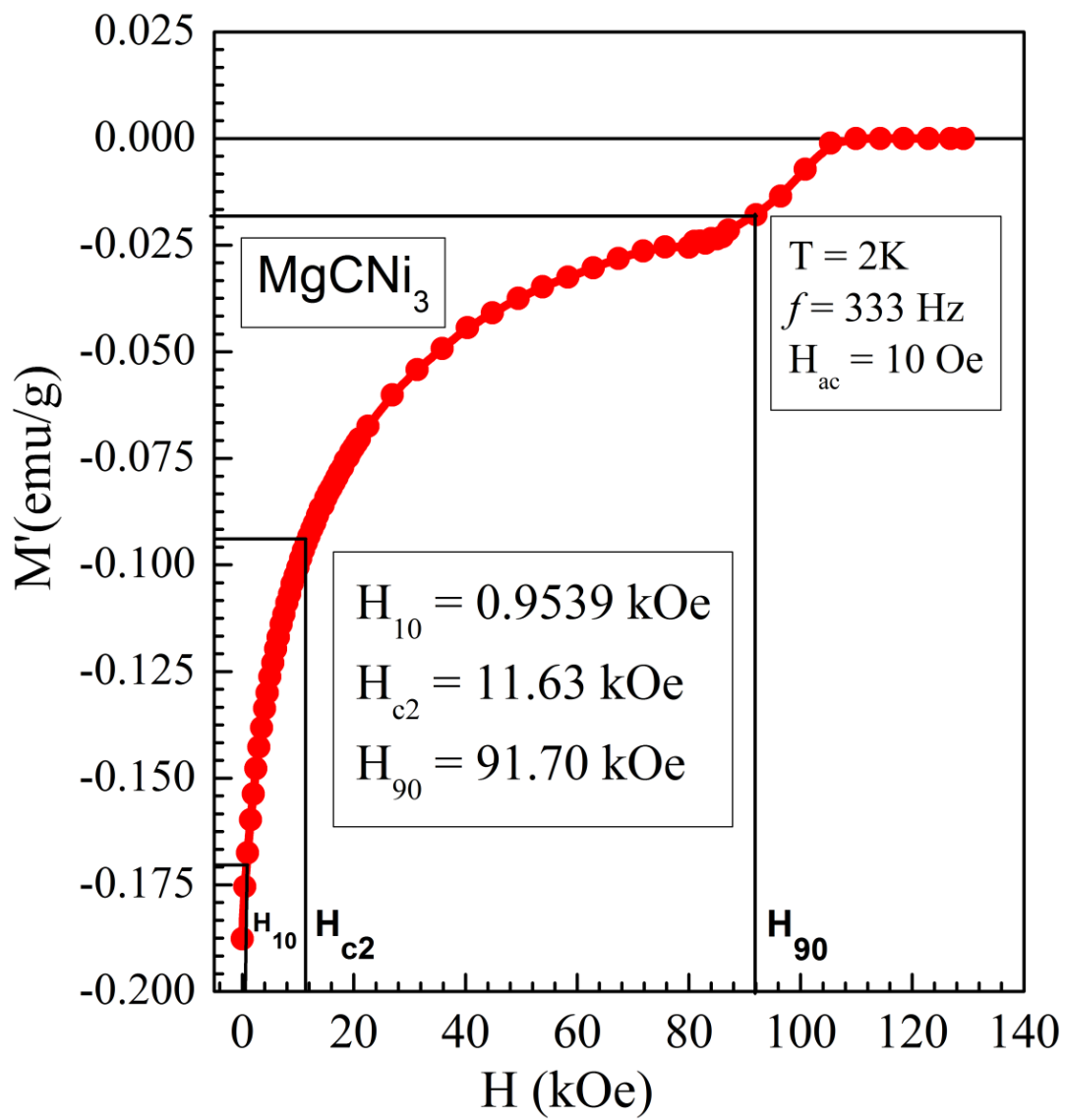

**Figure 7**

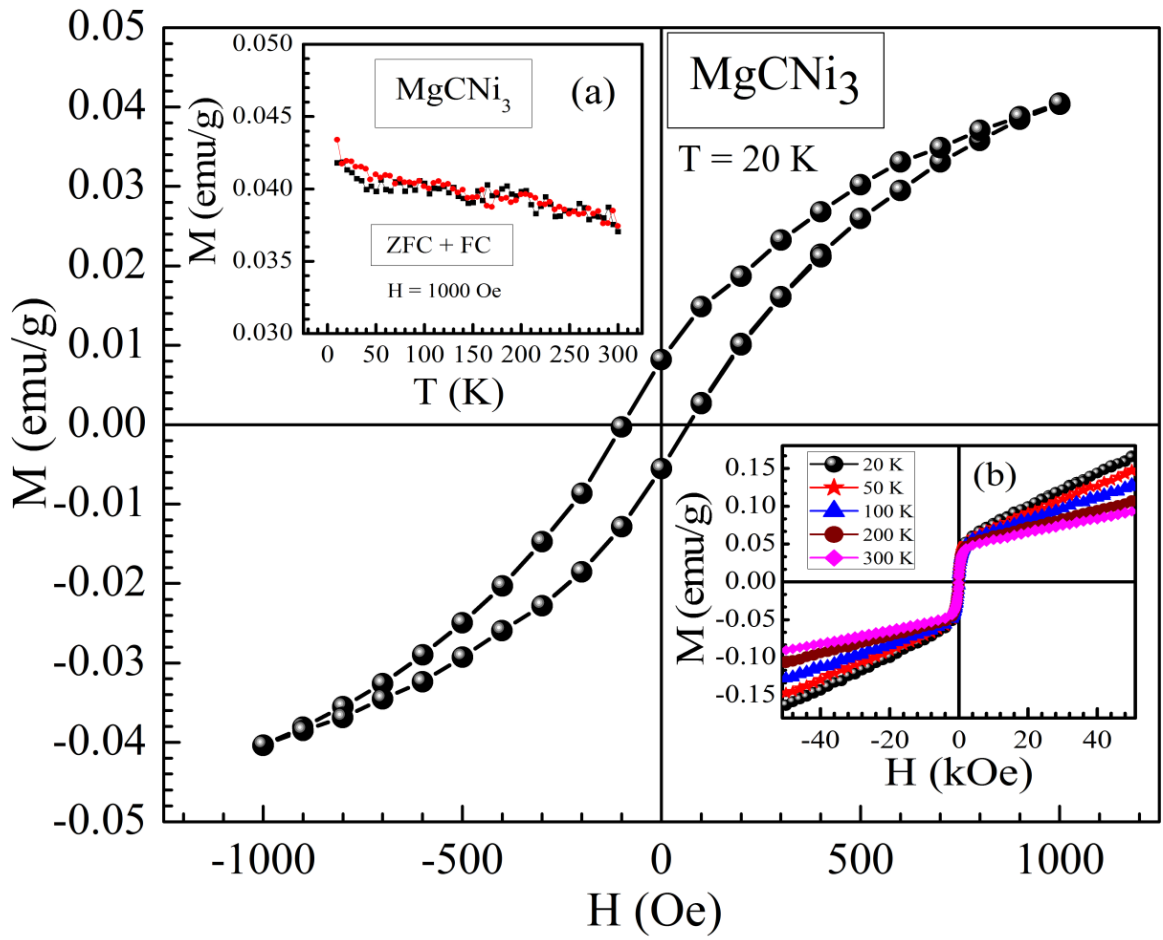

**Figure 8**

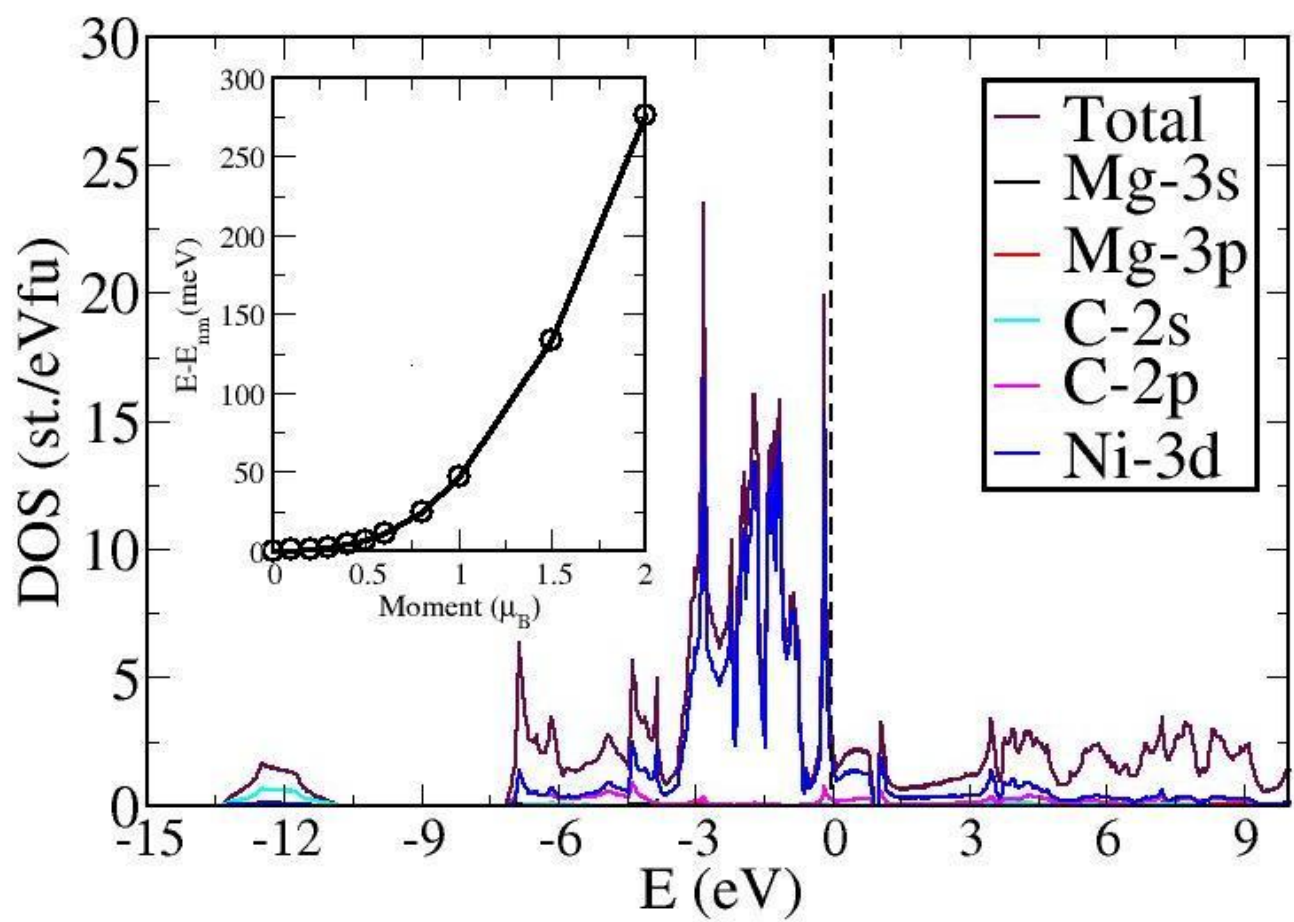

**Figure 9**

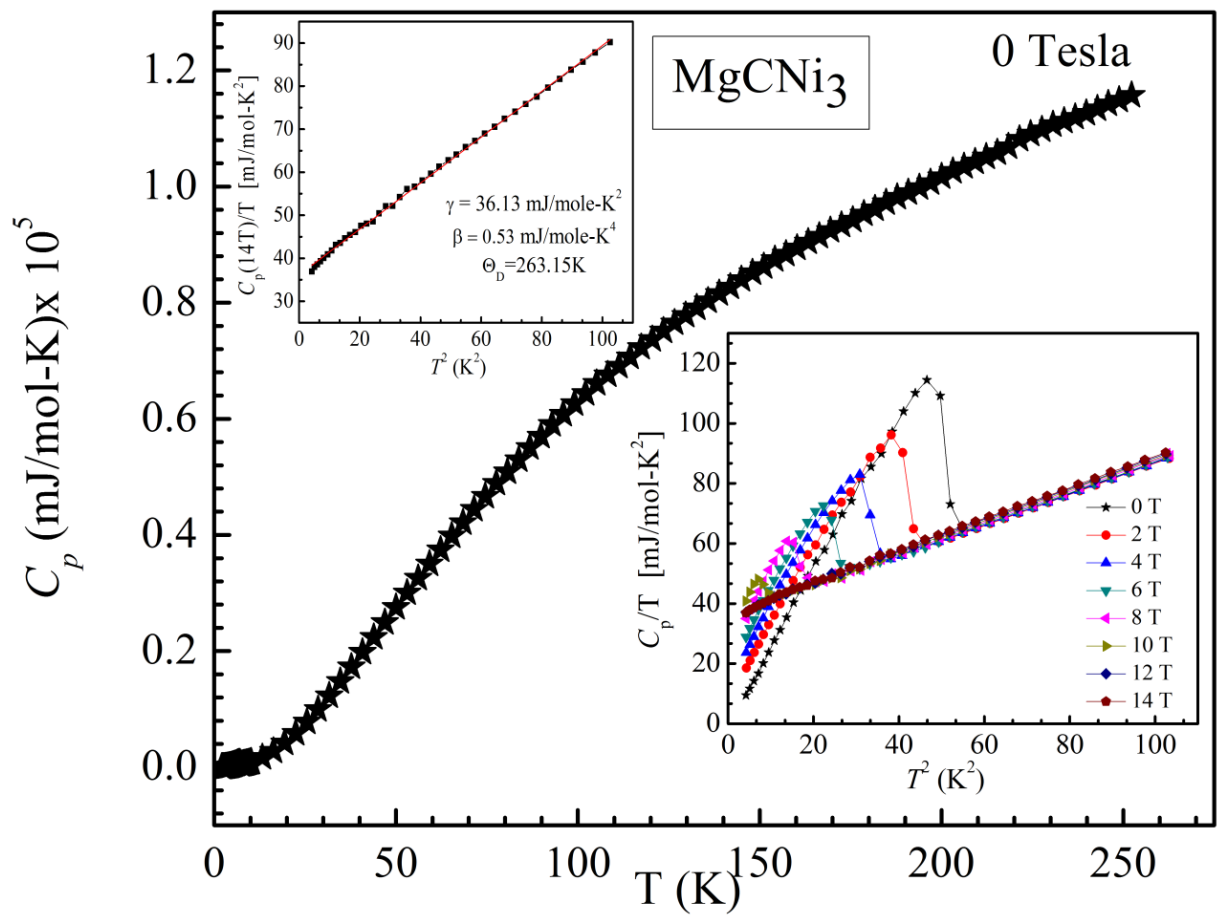

**Figure 10**

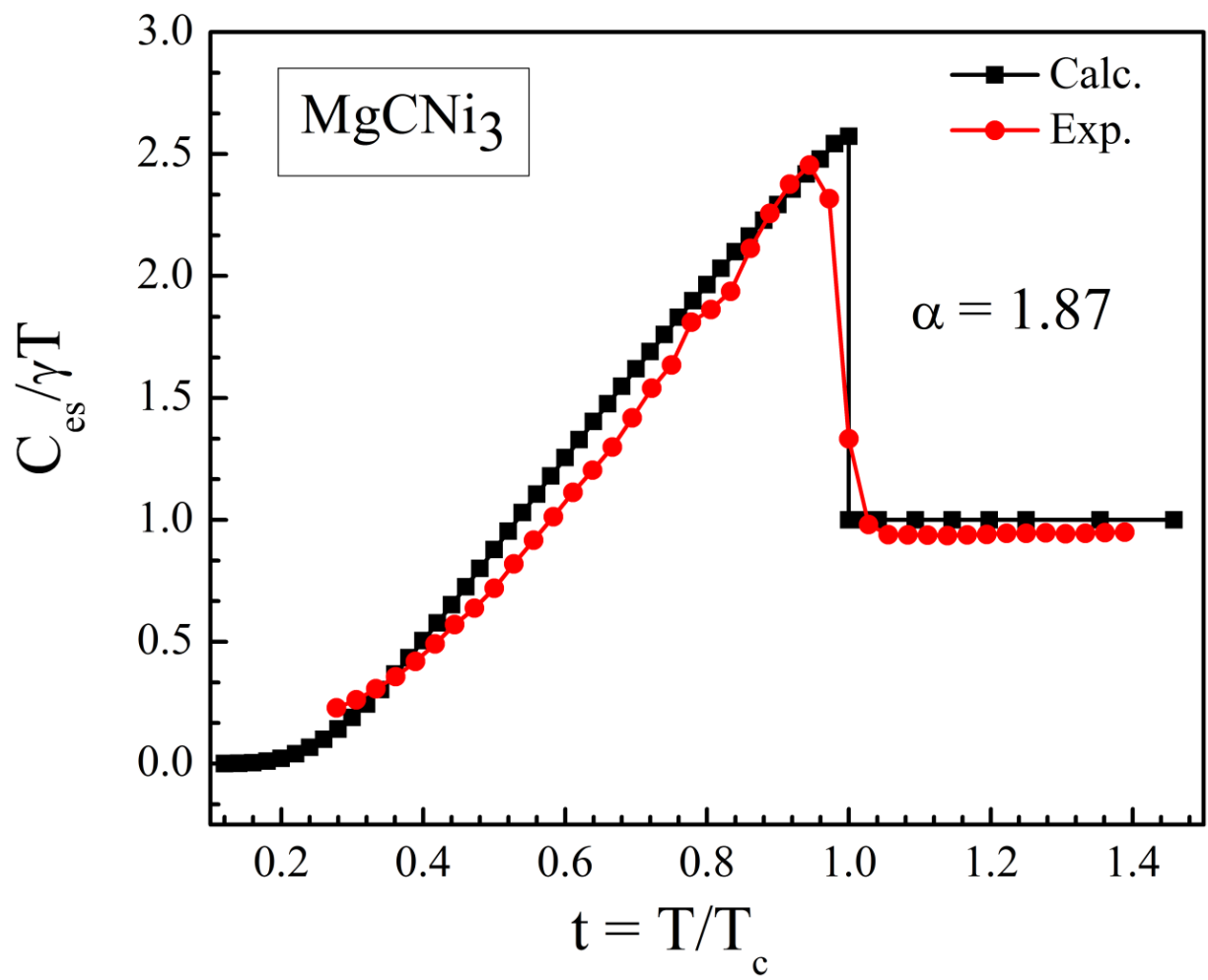